\documentclass{aa}  
\usepackage{graphicx}
\usepackage{color}
\usepackage{soul}
\usepackage{multirow}
\usepackage{booktabs}
\usepackage{subcaption}
\usepackage{natbib}
\usepackage{hyperref}
\usepackage{ulem}
\usepackage{txfonts}
\hypersetup{
colorlinks=true,
linkcolor=magenta,
citecolor=magenta,
filecolor=magenta, 
urlcolor=cyan,
}
\usepackage[switch]{lineno}

\begin{document} 

   \title{Radiation-driven stellar winds at the fast–slow transition: new hydrodynamic solutions}

   \author{M.C. Fernandez \inst{1}\thanks{Fellow of UNLP}, R.O.J. Venero \inst{1,2}, L.S. Cidale \inst{1,2}, I. Araya \inst{3} \and M. Cur\'e\inst{4}
          }

   \institute{Departamento de espectroscop\'ia, Facultad de Ciencias Astron\'omicas y Geof\'isicas, Universidad Nacional de La Plata, Paseo del Bosque S/N, BF1900FWA La Plata, Buenos Aires, Argentina
         \and
         Instituto de Astrof\'isica de La Plata, CCT La Plata, CONICET-UNLP, Paseo del Bosque S/N, BF1900FWA La Plata, Buenos Aires, Argentina
         \and
         Centro Multidisciplinario de F\'isica, Vicerrector\'ia de Investigaci\'on, Universidad Mayor, 8580745 Santiago, Chile
         \and
         Instituto de F\'isica y Astronom\'ia, Facultad de Ciencias, Universidad de Valpara\'iso, Av. Gran Bretaña 1111, Casilla 5030, Valpara\'iso, Chile
             }

   \date{Received January 9, 2026; accepted April 16, 2026}
 
  \abstract
   {Radiation-driven winds of massive stars can be described within the modified CAK theory, which parametrises the radiation force through three key quantities: $\alpha$, $\delta$, and $k$. Different combinations of these parameters, together with rotation, result in three types of stationary solutions, namely fast (or classical), $\delta$-slow, and $\Omega$-slow solutions.}
   {The primary objective of this work is to model radiation-driven winds inside the gap region between the fast and $\delta$-slow regimes, where stationary solutions have proven elusive. In addition, we compute synthetic line profiles of \ion{H}{i}, \ion{He}{i}, and \ion{Si}{iv} to illustrate the morphology of different wind regimes. }
   {We employ the time-dependent hydrodynamic code ZEUS-3D, capable of obtaining stationary solutions by progressing through an initial solution. Then we compute the line profiles solving the transfer equation for an expanding atmosphere, assuming spherical symmetry in the comoving frame, under non-local thermodynamic equilibrium (NLTE) conditions.}
   {We found new stationary solutions in the gap region, alongside their corresponding line profiles, for a typical B supergiant star model. In this model, the new solutions are stable, and some of them present a kink in the velocity profile at a fixed distance from the star, depending on the $\delta$ value. Perturbations in the wind ionisation may trigger transitions between different hydrodynamic regimes and offer a plausible explanation for structured and variable winds. A systematic investigation of these effects will be the subject of future work. Furthermore, we investigate the resulting line profiles from different hydrodynamic solutions and compare them with those predicted by a velocity profile given by a $\beta$-law using the same global wind parameters.}
    {}


\titlerunning{Radiation-driven stellar winds at the fast–slow transition: new hydrodynamic solutions}
\authorrunning{Fernandez et al.}
   \maketitle

\section{Introduction}

Winds from hot, luminous stars provide a natural laboratory for testing complex radiative‑transfer and hydrodynamic models. They play a central role in the evolution of massive stars, angular momentum loss, and the final stages of evolution (towards a neutron star or black hole). Moreover, such winds shape the stellar environment, injecting momentum, energy, and chemically processed material into the interstellar medium (ISM), thus regulating star formation and contributing to galactic chemical enrichment.

The winds of hot, massive stars are primarily driven by the transfer of momentum from the stellar radiation field to the plasma in the outer layers of the atmosphere, resulting in what are known as radiation-driven winds. These outflows were first described by \citet*{castorak1975}, whose work led to what is now known as the CAK theory. Subsequent extensions led to the modified CAK framework \citep[m-CAK,][]{Pauldrach1986,friendabbott1986}, which models the particle acceleration due to radiation pressure from a finite-size star. This theory also incorporates the effects of stellar rotation, applicable for rotation speeds up to 75\% of the critical rotation velocity. \citet{2023curearayareview} provided an updated overview of this theoretical framework.

A commonly adopted approach for modelling the winds of massive stars is to prescribe the velocity profile through the empirical "$\beta$-law" \citep{Pauldrach1986}. Theoretical expectations place $\beta$ in the range 0.5–1, yet in practice it is treated as a free parameter and pushed to much larger values ($\beta \approx 2 - 3$) to fit the H$\alpha$ line profile observed in B supergiant stars \citep[e.g.,][]{2008markovapuls,haucke2018}. However, such extrapolation falls outside the wind outflow regime where the $\beta$-law is physically justified, and it turns the wind parameters (mass-loss rate and terminal velocity) into adjustable quantities rather than predictions from a hydrodynamic model. A more physically motivated alternative is to derive the wind structure directly by solving the hydrodynamic equations, thereby ensuring that the resulting wind parameters are consistent with the underlying physics instead of being imposed through an assumed velocity law.

Although the m‑CAK theory reproduces most observed wind features in the early-type stars, several key questions remain unresolved: the non‑uniqueness of wind‑structure solutions when modelling line spectral features \citep{2024venero}, the lack of exact stationary solutions for the entire parameter space of the line-force multipliers \citep{venero2016}, the existence and character of the bi‑stability jump \citep[e.g.,][]{2008markovapuls,deburgos2024}, the interaction of slow and fast winds \citep{cure2011}, the coupling between wind variability and stellar pulsations \citep{cidale2023}, or the detailed properties of clumpy winds \citep[e.g.][]{rubiodiez2020}. Having exact hydrodynamic solutions is therefore essential to address outstanding problems, as hydrodynamics yields accurate velocity and density profiles for studying wind variability and instabilities, and supplying robust initial conditions for time‑dependent and multi‑D wind models.

Within the framework of the m-CAK theory, \citet{abbott1982} parametrised the radiative acceleration due to spectral lines using three key quantities: $k$, $\alpha$, and $\delta$. These parameters characterise, respectively, the effective number of contributing lines, the distribution of line strengths, and the sensitivity of the line-driving force to variations in the ionisation state. Depending on the value of the $\delta$ parameter, the hydrodynamic solutions may correspond either to fast (m-CAK–classical) regimes or to the so-called $\delta$-slow regimes \citep{cure2011}.  For instance, slowly rotating stars may develop either fast (for $\delta \lesssim 0.2$) or $\delta$-slow (for $\delta \gtrsim 0.28$) solutions, which differ markedly in their terminal wind velocities. However, for intermediate values of $\delta$, there exists a gap for a subset of line-force parameters where no stationary solutions connecting the fast and slow regimes had previously been found \citep{venero2016, 2024venero}. 

One of the primary objectives of this work is to model radiation-driven winds in 1D within the gap region. We employ the time-dependent hydrodynamic code ZEUS-3D  \citep{clarke1996, clarke2010}, which is capable of converging toward stationary solutions by iteratively advancing from an initial configuration. This approach allows us to probe and characterise the solutions emerging inside the gap region, thereby providing insights into the dynamics of radiation-driven winds in massive stars across this otherwise unexplored sector of parameter space. 
Once the hydrodynamic solutions within the gap region are obtained, we solve the radiative transfer equation under non-LTE conditions in the comoving frame, assuming spherical symmetry to compute synthetic line profiles. These profiles are then compared to quantify and assess the differences arising from the various wind solutions.

The paper is organised as follows: Section~\ref{section:2} reviews the stationary m-CAK theory, while Section~\ref{section:3} describes the numerical codes employed and the methodology followed. The resulting hydrodynamic solutions and the corresponding synthetic line profiles are presented in Section~\ref{section:4}. Section~\ref{section:5} presents a discussion, and Section~\ref{section:6} summarizes our main conclusions.

\section{The time-dependent m-CAK theory}\label{section:2}
This section provides a summary of the time-dependent m-CAK theory in the equatorial plane (one radial dimension) for radiation-driven winds \citep[e.g.,][]{Feldmeier1995, araya2018}. Following \citet{araya2018}, the outflow of gas from the outer layers of stars into space is mainly described by:
\begin{itemize}
    \item the mass-continuity equation,
    \begin{equation}
        \frac{\partial\rho}{\partial t} + \frac{1}{r^2} \frac{\partial}{\partial r} (\rho r^2 v) = 0 \textnormal{,}    \end{equation}
  \item and the momentum equation,
   \begin{equation}
     \frac{\partial v}{\partial t} + v \frac{\partial v}{\partial r} = -\frac{a^2}{\rho} \frac{\partial \rho}{\partial r} + g_{\textnormal{eff}} (r) + g_{\textnormal{rad}}^{L} (r,v,dv/dr) \textnormal{,}
     \label{Eq:2}
    \end{equation}
\end{itemize}
where $r$, $\rho(r,t)$, $v(r,t)$, $\partial v(r,t)/\partial r$, and $a$ are the radial coordinate, the mass density, the flow velocity, the velocity gradient, and the isothermal sound speed, respectively.

The effective gravity at the equatorial plane, $g_{\textnormal{eff}}$, is given by:
\begin{equation}\label{3}
    g_{\textnormal{eff}} (r) = - \frac{\rm{G} M_* (1 - \Gamma)}{r^2} \left( 1-\Omega^2 \frac{R_*}{r} \right)\textnormal{,}
\end{equation}
where  G is the gravitational constant, $M_*$ is the mass of the star, $R_*$ is its radius, and $\Omega = v_{rot}/v_{crit}$ is the ratio of the equatorial rotational velocity of the star to the critical velocity at the equator ($0<\Omega<1$). The parameter $\Gamma$ is the Eddington factor that accounts for the continuum radiative acceleration dominated by Thomson scattering. 

The acceleration due to the spectral lines, $g_{\textnormal{rad}}^{L}$, is given by:
\begin{equation}\label{4}
    g_{\textnormal{rad}}^{L} (r,v,dv/dr) = CF\, \frac{\Gamma\, \rm{G} M_*\,\it{k}}{r^2} \left( \frac{1}{\sigma_e v_{th}} \right)^{\alpha} \left( \frac{1}{\rho} \frac{\partial v}{\partial r} \right)^{\alpha} \left( \frac{n_e}{W(r)}\right)^{\delta}\textnormal{,}
\end{equation}
where $\sigma_e$ is the absorption coefficient due to Thomson scattering,  and $v_{th}$ is the thermal speed of the protons. In addition, $n_e$ is the electron density, and $CF(r,v,\partial v/\partial r)$ is the correction factor due to the finite stellar disk. In this work, we adopt the standard geometrical dilution factor given by
\begin{equation}
    W(r) = \frac{1}{2} \left(1-\sqrt{1-\frac{R_*^2}{r^2}} \right) \rm{,}
\end{equation}
which accounts for the finite angular extent of the stellar disk as seen from radius $r$. The finite-disk correction factor $CF$ is computed following the m-CAK formalism (e.g., \citealt{Pauldrach1986}), using
\begin{equation}
    CF = \frac{(1+\sigma)^{1+\alpha} - (1+\sigma \mu_*^2)^{1+\alpha}}{(1+\alpha)\sigma(1+\sigma)^{\alpha}(1-\mu_*^2)} \rm{,}
\end{equation}
where $\mu_* = \sqrt{1-R_*^2/r^2}$ and $\sigma = d\ln\,v/d\ln\,r -1$.

An alternative description of the radiative line force was introduced by \citet{1995gayley}, who reformulated the CAK parametrisation in terms of the dimensionless line-strength parameter $\overline{Q}$ and the cut-off parameter $Q_0$, instead of the line-force parameter $k$. In this formalism, the line acceleration $g_{\rm{rad}}^L$ takes the form
\begin{equation}
    g_{\rm{rad}}^L (r,v,dv/dr) = CF \, \frac{\Gamma G M_* \overline{Q}}{(1-\alpha) r^2} \left(\frac{1}{Q_0 \sigma_e c}  \right)^{\alpha} \left( \frac{1}{\rho}\frac{\partial v}{\partial r} \right)^{\alpha}\left( \frac{n_e}{W(r)}\right)^{\delta} \rm{.}
\end{equation}
The relation between the Gayley parameters and the standard CAK line-force parameters can be written as
\begin{equation}
    k = \frac{1}{1-\alpha} \left( \frac{v_{\rm{th}}}{c} \right)^{\alpha} \overline{Q} Q_0^{-\alpha} \rm{.}
\end{equation}
We adopt $Q_0 \approx \overline{Q}$, treating both parameters as representative of the same underlying line-strength distribution. This allows us to derive $\overline{Q}$ from the adopted values of $k$ and $\alpha$.

The line-force parameters $\alpha$, $k$, and $\delta$ define the analytical parametrisation of the line-driving acceleration in the m-CAK theory. Especially relevant for our work is the parameter $\delta$, introduced by \citet{abbott1982}, which accounts for changes in ionisation throughout the wind. 

The m-CAK theory delivers three physical stationary solutions \citep{2023curearayareview} depending on $\delta$ and the rotation rate, $\Omega$: 
\begin{itemize}
    \item The fast solution has a slow rotation rate ($\Omega \lesssim 0.75$) and almost no changes in the ionisation of the wind with distance (i.e., $\delta \lesssim 0.2$). This is known as the classical m-CAK solution \citep{Pauldrach1986, friendabbott1986}, and predicts a high terminal velocity.
    
    \item The $\Omega$-slow solution corresponds to models with high rotation rate ($\Omega \gtrsim 0.75$) and yields to a low terminal velocity  \citep{cure2004}.
    
    \item The $\delta$-slow solution results from high values of $\delta$ (usually $\delta \gtrsim 0.28$) and leads to low values for the terminal velocity \citep{cure2011}.
\end{itemize}

The fast and $\delta$-slow solutions are separated by a region in the $\delta$-parameter space, or "gap" \citep{venero2016}, where no steady hydrodynamic solution could be found using the Hydwind code \citep{cure2004}.  
As the Hydwind code solves the hydrodynamic equations under spherical symmetry and in a stationary state, both fast and $\delta$-slow solutions can be achieved without significant convergence issues for most radiation force parameters. However, when $\delta$ values approach the critical region (or gap), the solutions fail to converge. This gap interrupts the smooth transition of hydrodynamic regimes along the domain of solutions. To find the solution within the gap, we employ the time-dependent code ZEUS-3D \citep{clarke1996, clarke2010}, following the scheme developed by \citet{araya2018}.

\section{Methodology}\label{section:3}

\subsection{Time-dependent hydrodynamic models}
The code ZEUS and its subsequent extensions, ZEUS-2D and ZEUS-3D, have been among the most influential numerical tools in computational astrophysics since their introduction by \citet*{1992stonenorman}, and \citet*{1992stonemihalasnorman}. Its robust algorithms for solving the equations of hydrodynamics and magnetohydrodynamics, together with its flexibility for multidimensional simulations, enabled pioneering studies of numerous astrophysical flows. The code has been widely applied to problems such as accretion disks and magnetorotational turbulence, protostellar jets and outflows, turbulence in the interstellar medium, star formation, and the hydrodynamics of stellar winds \citep[e.g.,][]{1995hawley,1997ouyed,2003fujita,2012barai,venero2016,araya2018}.

The ZEUS-3D code employs an Eulerian finite-difference, operator-split scheme with 
consistent advection for transport \citep{clarke1996, clarke2010}. The line acceleration was implemented following the parametrisation of \citet{1995gayley}, as adapted by \citet{araya2018}. The radial velocity gradient is computed using a first-order forward finite difference, and the resulting line acceleration is then added explicitly as a source term in the momentum equation.

At the inner boundary, we impose the base density, while the inflow velocity is allowed to float and is obtained by linearly extrapolating the velocity from the two innermost grid zones of the computational domain. In practice, the base density is adjusted in order to obtain a stable outflow and allow the hydrodynamic solution to converge towards a stationary state. At the outer boundary, we impose a standard outflow (zero-gradient) condition. No fixed terminal velocity is enforced. The simulations were performed in one radial dimension using a grid of 500 points extending from the stellar radius up to 100~R$_*$. The mesh is non-uniform, employing logarithmic radial coordinates to enhance the grid resolution near the stellar surface, where strong gradients are expected, following the prescription of \citet{araya2018}. 

To initialise ZEUS‑3D, we need to supply converged stationary solutions that are obtained from the Hydwind code, providing the stellar and line-force parameters. Our procedure is as follows. Utilising the Hydwind code, we first seek the position of the gap in the $\delta$-parameter space. Starting from a low $\delta$ value in the fast wind regime, we increase $\delta$ until the highest value that achieves a converged solution, marking the lower boundary of the gap, hereafter referred to as $\delta_{\textnormal{min}}$. Then, a $\delta$ value corresponding to the slow regime is chosen (for example, $\delta = 0.4$), and this value is gradually decreased. The final $\delta$ value at which a converged solution is achieved defines the upper boundary of the gap, denoted as $\delta_{\textnormal{max}}$. Thus, for a given set of stellar and line-force parameters ($\alpha$, $k$), together with the rotation rate $\Omega$, the gap is defined as the interval $\delta_{\textnormal{min}} < \delta < \delta_{\textnormal{max}}$.

ZEUS-3D solves the time-dependent hydrodynamic equations and, unlike Hydwind, does not impose the critical-point condition to construct a stationary solution. Instead, the equations are integrated numerically from the stellar surface outwards, and the system evolves in time until a steady state is reached. In this approach, the final solution is determined solely by the adopted stellar and line-force parameters. 

\begin{figure}
\centering
\includegraphics[width=\hsize]{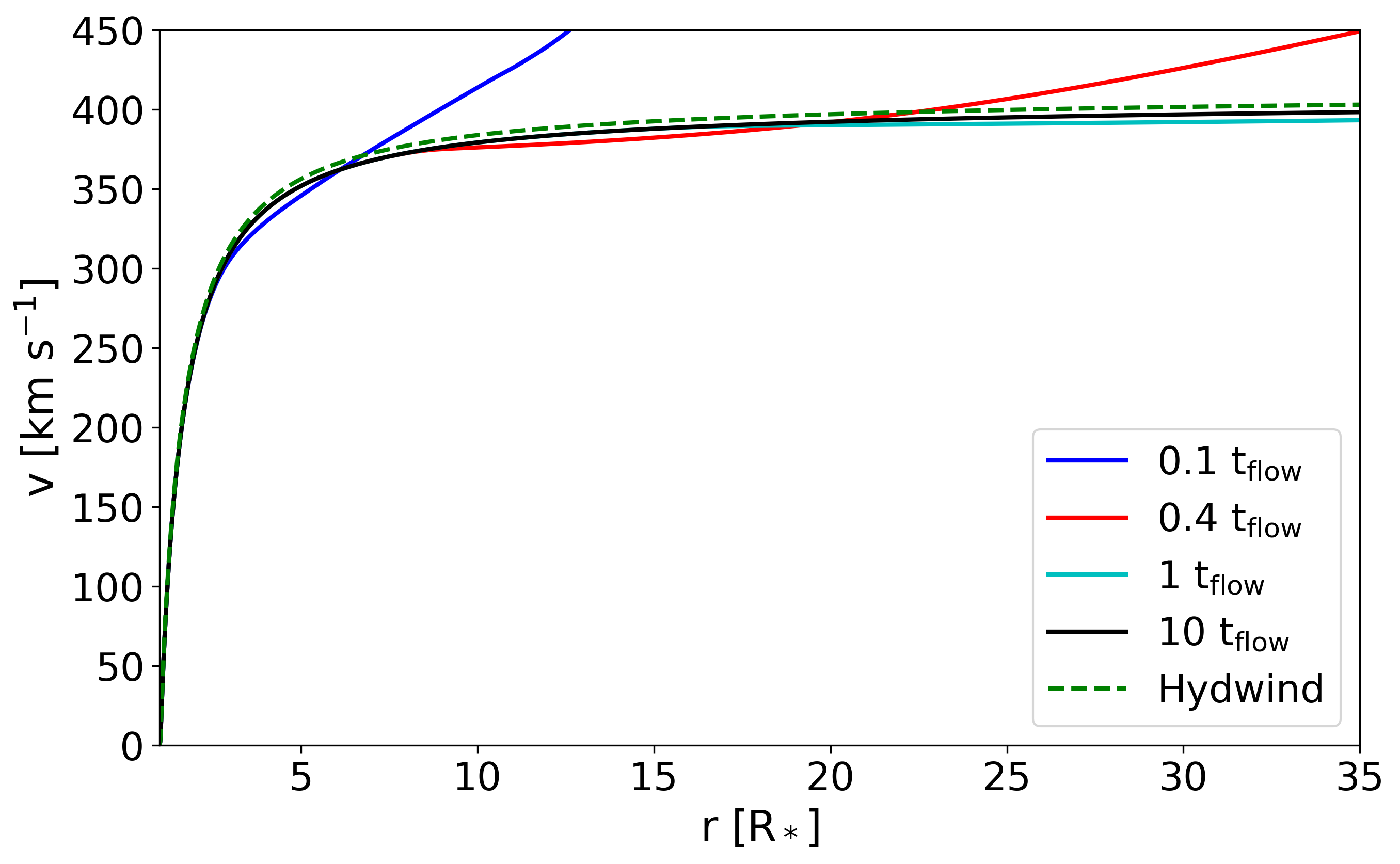}
\includegraphics[width=\hsize]{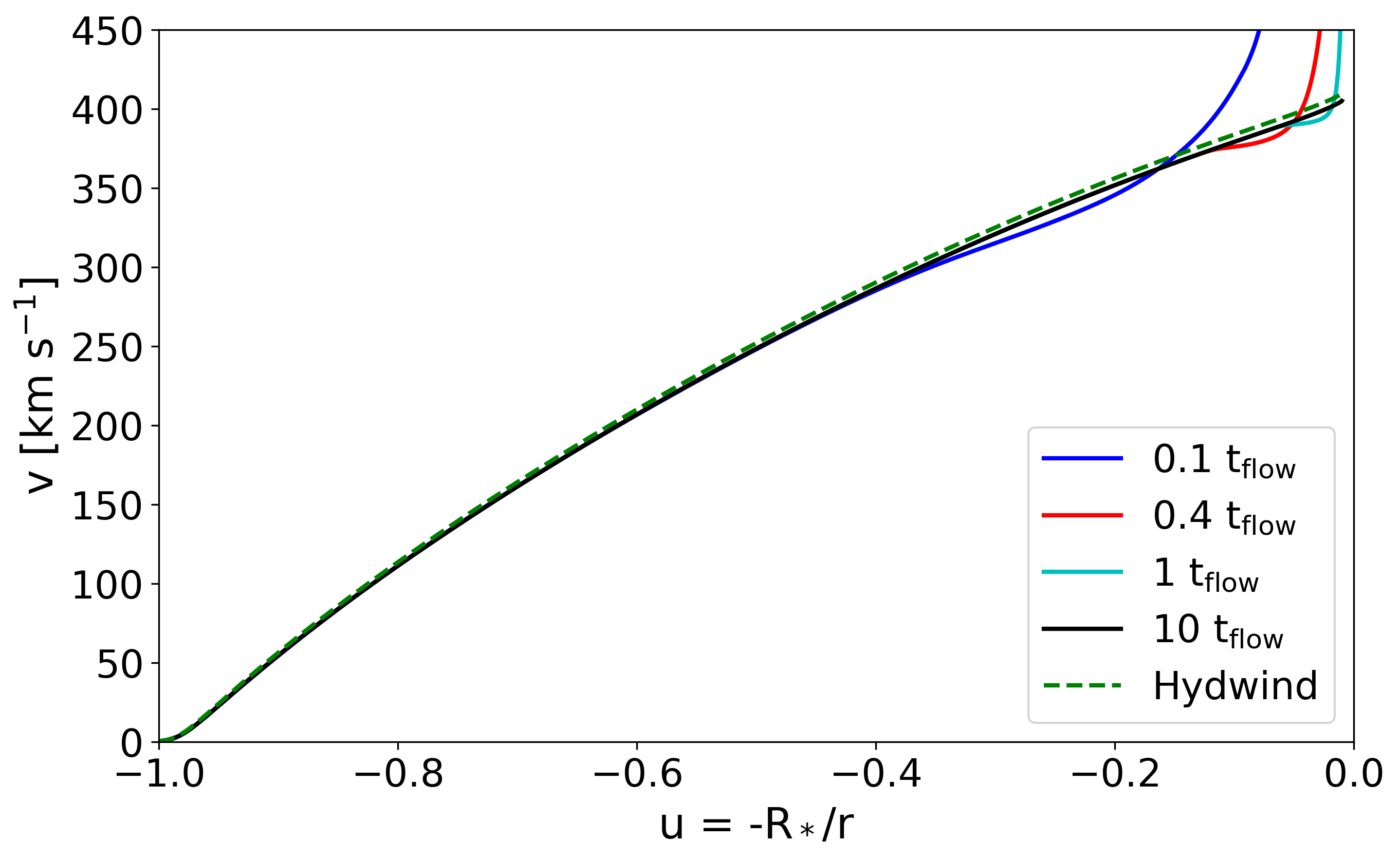}
    \caption{Temporal evolution obtained with ZEUS-3D (solid lines) from an initial solution (dashed line). Times of the different velocity laws are given in units of the flow crossing time.}
    \label{fig:zeus-temp}
\end{figure}

To illustrate the relaxation towards the stationary regime, Fig.~\ref{fig:zeus-temp} shows the temporal evolution of the velocity field $v(r)$ computed with ZEUS-3D. The curves correspond to different times expressed in units of the flow crossing time, defined as t$_{\rm{flow}} = \rm{R}_{max} / \rm{v}_{\infty}$, where R$_{\rm{max}}$ is the outer radius of the computational domain and v$_{\infty}$ is the terminal velocity of the wind. For the model shown here, characterised by T$_{\textnormal{eff}} = 18\,000$~K, $\log\,g = 2.5$, R$_* = 23$~R$_{\odot}$, $\Omega = 0.27$, $\alpha= 0.515$, $k= 0.104$, $\delta= 0.1$, and $\overline{Q} = 67.25$, we adopt a base wind density of $\rho_0 = 10^{-12}$ g cm$^{-3}$. For consistency in the comparison, the stationary solution computed with Hydwind was obtained using the same base density, instead of the usual condition in which the stellar radius is defined at the optical depth $\tau = 2/3$. For this model, the flow crossing time is approximately t$_{\rm{flow}} \approx 4 \times 10^6$~s. The figure displays the velocity profile at $0.1$, $0.4$, $1$, and $10 ~\rm{t}_{\rm{flow}}$. At early times, the inner wind regions already approach their stationary structure, while the outer layers are still evolving toward a steady state. In particular, within approximately $r \lesssim 2R_*$, the solution is already close to the final profile at $0.1~\rm{t}_{\rm{flow}}$. As the simulation proceeds, the outer wind gradually relaxes, and by $\rm{t} \approx t_{\rm{flow}}$ the entire domain has essentially reached the stationary configuration. The simulation is nevertheless continued up to $10~t_{\rm{flow}}$, which largely exceeds the time required for convergence. For comparison, Fig.~\ref{fig:zeus-temp} also includes the stationary solution computed with Hydwind for the same stellar and line-force parameters. The time-dependent solution obtained with ZEUS-3D clearly converges towards the same velocity profile, confirming the consistency between both approaches.

In some cases, transient features may appear during the relaxation phase. For instance, when the initial condition corresponds to a wind solution belonging to a different regime, the intermediate velocity profiles can display temporary distortions while the flow readjusts to the correct stationary solution \citep[see discussion in][]{araya2018}. These features disappear once the entire domain has relaxed and do not affect the final converged state, provided that the simulation is evolved for a sufficiently long time.

To verify that the final solution is independent of the initial velocity profile, we used different initial conditions, including solutions from different wind regimes. In all cases, the simulations converge towards the same stationary solution determined by the input stellar and line-force parameters. This property is particularly relevant for exploring the gap region in parameter space, where no Hydwind solution is available.

The time integration in ZEUS-3D uses explicit time stepping with a Courant–Friedrichs–Lewy coefficient of 0.5. In practice, the simulations are evolved well beyond the time required for convergence. Specifically, the equations are integrated from \mbox{$\rm{t}=0$} up to \mbox{$\rm{t}_{\rm max} = 9 \times 10^7$~s}, using a fixed timestep of \mbox{$\Delta \rm{t} = 9 \times 10^4$~s}, resulting in $1000$ temporal steps in total. Rather than identifying the exact moment at which the steady state is reached, we simply evolve the system for a sufficiently long time and adopt the final snapshot as the stationary solution. Convergence is further verified by checking that the mass-loss rate becomes constant throughout the computational domain.

\subsection{Radiative transfer calculations for spectral lines}

The atmospheric structure of the star consists of a photospheric temperature and density stratification smoothly connected to an isothermal stellar wind. The transition between the two regions is implemented through a continuous interpolation. Subsequently, following the hydrodynamic calculations described above, we computed synthetic spectra to assess which stationary solutions most accurately reproduce the observed line profiles. To this end, we solved the radiative transfer equation in an expanding atmosphere under non-local thermodynamic equilibrium (NLTE), assuming spherical symmetry and stationarity in the comoving frame \citep[following][]{mihalaskunasz1978}. The radiative transfer and statistical equilibrium equations are iterated simultaneously until convergence is achieved.

These calculations were performed with the MULITAS code ({\it MUlti LIne Transfer for Active Stars}), which includes a multilevel radiative transfer treatment for \ion{He}{ii} and \ion{Mg}{ii} ions in the comoving frame \citep{venerocidaleringuelet2000, cidale1998}, relevant for massive-star winds. The MULITAS code was also adapted by \citet{cidale1993} to model the \ion{Si}{iv} atom (six bound levels plus continuum), allowing us to compute its ultraviolet (UV) resonance doublet.

In this work, we have extended the MULITAS code to include a multilevel \ion{He}{i} atom, incorporating both singlet and triplet systems. A total of 23 bound levels plus a continuum were implemented. Energy levels and oscillator strengths for radiatively permitted transitions were obtained from the NIST database \citep{NIST}. We also incorporated the corresponding collisional rates (bound–bound and bound–free) and radiative cross-sections for the relevant levels \citep[cf.,][]{tesisrene}. This extension enables the computation of \ion{He}{i} lines in both the optical and infrared (IR), complementing the UV diagnostics provided by \ion{Si}{iv}.

To model the hydrogen lines, we employed the Appel code \citep{catalakunasz1987, cidaleringuelet1993}, which solves the radiative transfer equation in the comoving frame for continua and selected spectral lines under the same physical assumptions of sphericity and stationarity. The code Appel provides the H line profiles corresponding to the hydrodynamic velocity fields obtained from ZEUS-3D, and is used in the same manner as MULITAS for the calculation of \ion{Si}{iv} and \ion{He}{i} lines. 

Together, these radiative transfer calculations enable us to test the hydrodynamic models across the UV, optical, and IR spectral ranges, using a consistent set of stationary wind structures. This approach also allows us to assess which wind solution produces the most distinct spectral signatures.\\

\section{Results}\label{section:4}

To model a B supergiant star, we adopt the T19 model from \citet{venero2016} as a representative case to explore the new set of hydrodynamic solutions. The parameters of this model are: T$_{\textnormal{eff}} = 19~000$~K, $\log\,g = 2.5$, R$_* = 40$~R$_{\odot}$, $\alpha = 0.5$, $k = 0.32$, and $\overline{Q} = 433.1$. In addition, several values of the rotation rate $\Omega$ and the line-force parameter $\delta$ were considered to sample the full family of solutions, including the fast regime, the $\delta$-slow regime, and the transition from fast to $\delta$-slow solutions (i.e., in the gap region). 

\subsection{Hydrodynamic solutions}

\begin{figure*}[ht!]
    \centering
    \includegraphics[width=0.9\textwidth]{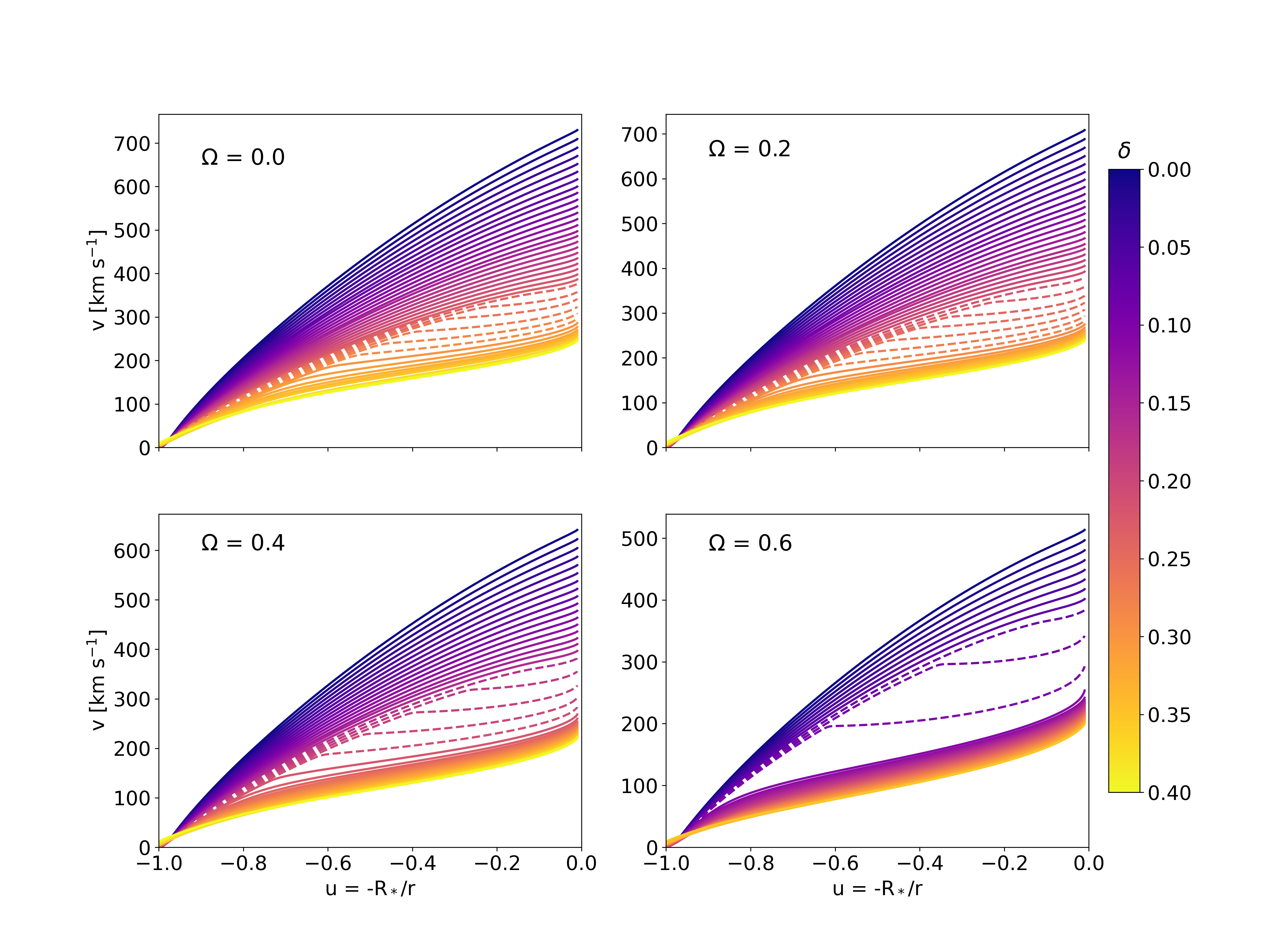}
    \caption{Hydrodynamic solutions for T19 model (T$_{\rm{eff}} = 19\,000$~K), adopting values from $0 < \delta < 0.4$ and $\Delta \delta = 0.01$, using different rotational rates, obtained with ZEUS-3D code. Dashed lines represents the solutions in the gap region.}
    \label{fig:solutions}
\end{figure*}

Figure~\ref{fig:solutions} displays a sequence of hydrodynamic solutions for model T19, computed with the ZEUS-3D code, considering rotation rates of $\Omega =$~0.0, 0.2, 0.4, and 0.6. In the plot, each curve corresponds to a different value of $\delta$, ranging from 0.0 to 0.4 with a step of 0.01. The solutions with the highest terminal velocities correspond to the fast regime, while those with the lowest terminal velocities correspond to the $\delta$-slow regime. In between, a new family of solutions emerges; these are the new solutions reported in this work (dashed lines in the plot).

Despite the uniform sampling in $\delta$, the transition-region solutions exhibit a noticeably wider spacing compared to both the fast and the $\delta$-slow regimes. This effect becomes more evident as the rotation rate increases.
The widening of the spacing reflects the rapid structural changes occurring in the wind as the flow transitions between the two classical regimes, highlighting the sensitive dependence of the solution on the ionisation parameter in this interval.

These new solutions form a continuous transition between the fast and $\delta$-slow regimes. Near the photosphere, they behave similarly to the fast solutions, but as the distance from the star increases, they gradually resemble the $\delta$-slow wind regime. This transition introduces a change in the slope $dv/du$ (where $\rm{u}=-\,\rm{R}_*/\rm{r}$), which produces a distinct kink in the velocity profile. When comparing different rotation rates, this kink becomes steeper as the stellar rotation increases. However, at higher values of $\Omega$, the situation becomes more critical: the flow approaches conditions where the $\Omega$-slow regime is expected to appear, which lies outside the scope of this work. In addition, B supergiants are generally slow rotators, so such high rotation rates are not expected to be representative of these stars.

A key result is that these kinks are stable structures of the wind, preserved even after long temporal integrations. As the simulations evolve for longer times, the solutions retain their shape, including the kink, confirming that they correspond to stationary states. Moreover, as $\delta$ increases, the kink appears progressively farther from the stellar surface. Therefore, if the ionisation conditions were to vary (see Section~\ref{subsection:5.3} for a discussion), effectively modifying the value of $\delta$, the kink may propagate outward through the stellar wind. This behaviour could be related to wind variability and to the formation and evolution of discrete absorption components, as reported in OB stars \citep[e.g.,][]{kaper1994, kaper1999}.

\begin{figure*}
    \centering
        \includegraphics[width=0.9\textwidth]{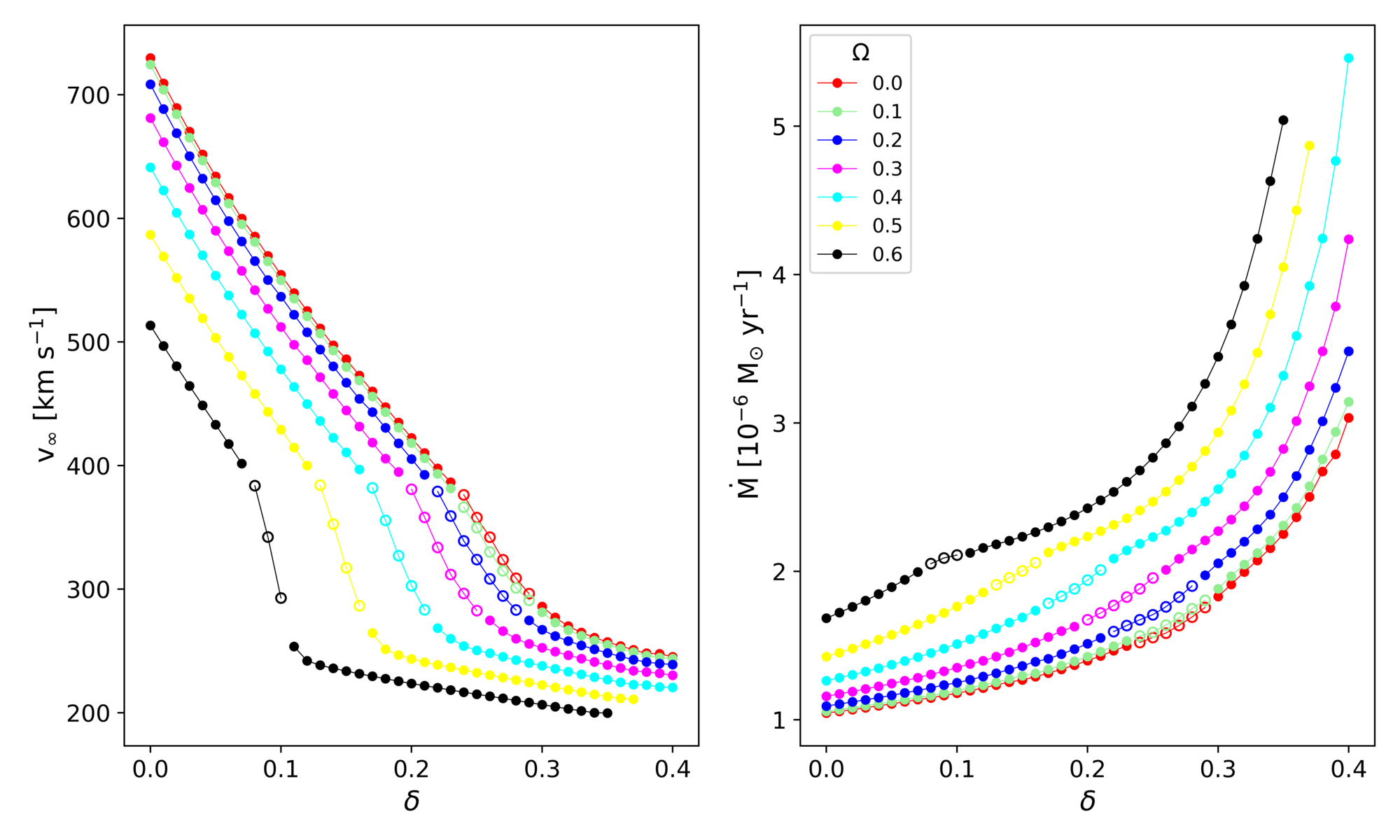}
    \caption{Terminal velocity (left panel) and mass-loss rate (right panel) as functions of the $\delta$ parameter for the hydrodynamic solutions obtained with ZEUS-3D for the T19 model, for several rotation rates. Filled symbols correspond to fast and $\delta$-slow solutions, while open circles denote the solutions found within the transition region.}
    \label{fig:vinf-mdot}
\end{figure*}

Figure~\ref{fig:vinf-mdot} presents the terminal velocity and mass-loss rate as functions of the $\delta$ parameter for all rotation rates considered (from 0.0 to 0.6, in steps of 0.1). Filled symbols mark the fast and $\delta$-slow solutions, while open circles denote the intermediate solutions found within the transition region. The left panel shows the variation of the terminal velocity with $\delta$, while the right panel displays the corresponding mass-loss rates. The result predicts that a decrease in terminal velocity is related to an increase in mass-loss rate. This figure extends the results of \citet{venero2016}, now fully covering the region where no stationary solutions were previously found. The ZEUS-3D simulations fill this gap, providing a continuous sequence of hydrodynamic solutions that smoothly bridge the fast and $\delta$-slow regimes. The previously missing intermediate region is now fully resolved, revealing that the transition between both regimes is continuous, both in the structure of the flow (through the velocity profiles) and in the resulting global wind parameters (v$_\infty$ and $\dot M$).

Furthermore, Table~\ref{table:1} lists the terminal velocities and mass-loss rates predicted by both Hydwind and ZEUS-3D for the different $\delta$ values, taking $\Omega = 0.2$ as a representative case. While Hydwind shows no solutions within the interval $0.21 < \delta < 0.29$, ZEUS-3D successfully finds stationary solutions for all $\delta$ values. Although we have explored only a single model, it is important to emphasize that both codes yield similar terminal velocities and mass-loss rates in the fast regime. In contrast, in the $\delta$-slow regime, the predicted mass-loss rates differ by up to a factor of two, with ZEUS-3D systematically yielding lower values.

\subsection{Synthetic profiles}

\begin{figure*}[ht!]
\centering
    \includegraphics[width=\textwidth]{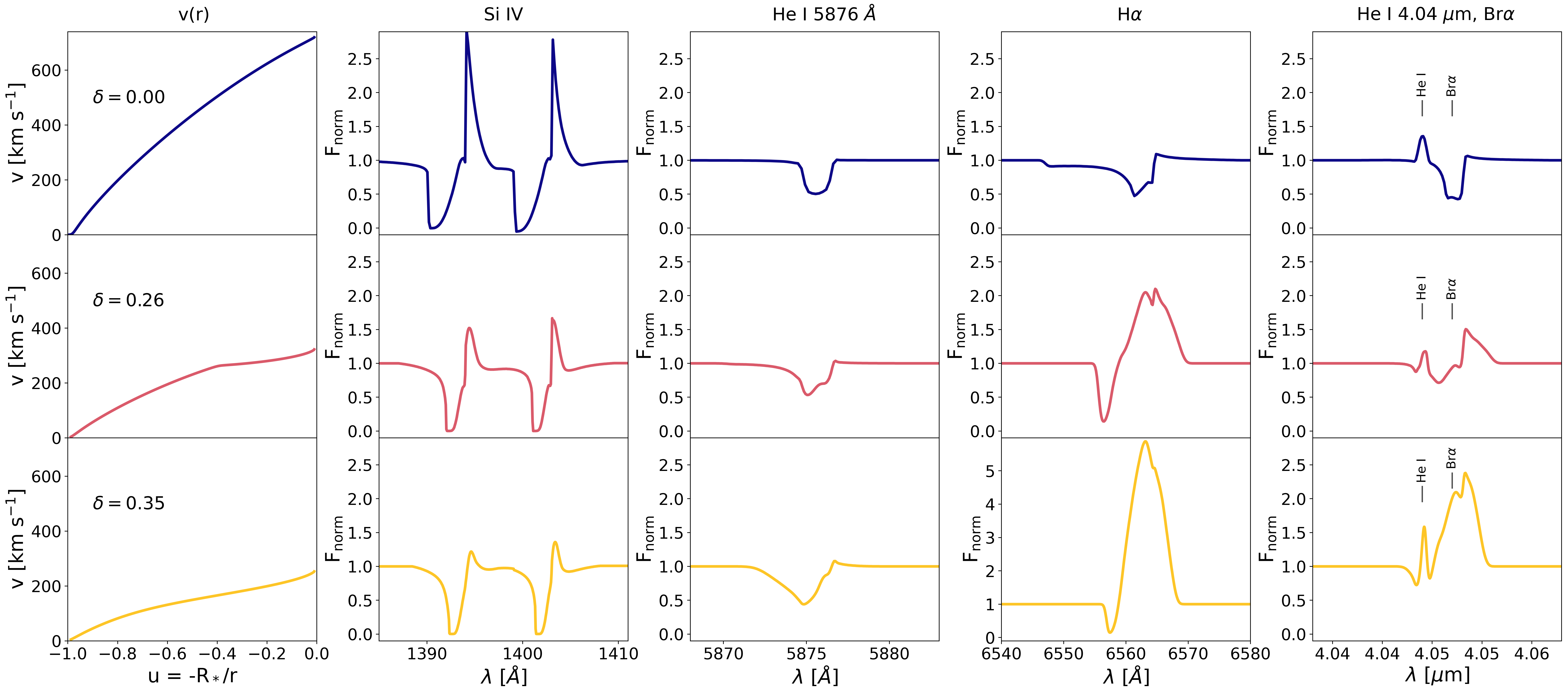}
    \caption{Synthetic line profiles computed for the three hydrodynamic wind regimes using the T19 model: fast solution ($\delta = 0$, top row), transition solution ($\delta = 0.26$, middle row), and $\delta$-slow solution ($\delta = 0.35$, bottom row). The first column shows the hydrodynamic solution obtained with ZEUS-3D, followed by the synthetic profiles for \ion{Si}{iv} (second column), \ion{He}{i} 5876~$\AA$ (third column), H$\alpha$ (fourth column), and the \ion{He}{i}~+~Br$  \alpha$ blend around 4.05~$\mu$m (fifth column). The fast solution corresponds to $\dot{M} = 1.06 \times 10^{-6}$~M$_{\odot}$~yr$^{-1}$ and v$_{\infty} = 720$~km~s$^{-1}$, the solution in the transition region to $\dot{M} = 1.67 \times 10^{-6}$~M$_{\odot}$~yr$^{-1}$ and v$_{\infty} = 322$~km~s$^{-1}$, and the $\delta$-slow solution to $\dot{M} = 2.35 \times 10^{-6}$~M$_{\odot}$~yr$^{-1}$ and v$_{\infty} = 253$~km~s$^{-1}$.
              }
    \label{fig:profiles}
\end{figure*}

Figure~\ref{fig:profiles} presents the synthetic line profiles computed for three representative hydrodynamic solutions of model T19: a fast solution ($\delta = 0.00$), a transition-region solution ($\delta = 0.26$), and a $\delta$-slow solution ($\delta = 0.35$), all computed for $\Omega = 0$. The first column shows the corresponding velocity profiles, while the remaining columns display the synthetic profiles for \ion{Si}{iv} doublet $\lambda\lambda$1394, 1403\,\AA, \ion{He}{i} 5876~$\AA$, H$\alpha$, and the \ion{He}{i} line + Br$\alpha$ blend near 4.04~$\mu$m.

Across all wavelengths, the profiles form a coherent morphological sequence that reflects the continuity of the underlying hydrodynamic structure. The transition-region solution consistently shows intermediate behaviour, matching its position between the fast and $\delta$-slow regimes.

The \ion{Si}{iv} doublet displays the expected P~Cygni morphology in all three cases. As the terminal velocity decreases from the fast to the $\delta$-slow solution, the width of the blue absorption trough becomes progressively narrower, while the emission component also weakens along the same sequence, even when the mass loss rate is increasing from $1.06\times10^{-6}$ to $2.35\times10^{-6}\rm{M}_{\odot}yr^{-1}$.

The \ion{He}{i}~$\lambda$5876~$\AA$ line begins as a pure absorption feature in the fast solution and gradually develops a weak red emission wing for the $\delta$-slow solution. Although the variation is modest, the intermediate profile is clearly bracketed by the fast and $\delta$-slow solutions, providing a clean spectroscopic counterpart to the underlying velocity structure and mass-loss rate.

The H$\alpha$ line shows a much stronger response to changes in the wind structure. The fast solution remains dominated by absorption, while the transition model already forms a clear P~Cygni profile. In the $\delta$-slow case, the emission lobe becomes very strong, consistent with the substantial increase in mass-loss rate (from $1.06\times10^{-6}$ to $2.35\times10^{-6}\rm{M}_{\odot}yr^{-1}$). 

In the infrared, both the \ion{He}{i} $\lambda$4.04~$\mu$m line and Br$\alpha$ respond sensitively to the structural changes induced by increasing $\delta$. The \ion{He}{i} line evolves from pure emission to a well-defined P~Cygni shape in the $\delta$-slow regime. The Br$\alpha$ line evolves from absorption to a strong P~Cygni profile, with the emission peak dominating in the slow-wind case. When combined, both lines appear largely in emission for the $\delta$-slow solution, as the \ion{He}{i} emission masks the blue absorption component of the Br$\alpha$ P~Cygni profile. For instance, both lines were reported in emission in the B supergiant \object{55 Cygni} \citep{cidale2023}.

Overall, all diagnostics consistently place the transition-region solution between the fast and $\delta$-slow regimes, reinforcing the continuity of the physical sequence revealed by the ZEUS-3D models. This behaviour indicates that the transition solution naturally connects both regimes within a continuous family of wind solutions.

\section{Discussion}\label{section:5}

\subsection{Why the Hydwind code fails to find solutions within the gap}

Under particular values of the radiation-force parameters, especially for $\delta$ values within the interval ($\delta_{\rm{min}}$, $\delta_{\rm{max}}$), the Hydwind code is unable to obtain a physically consistent wind solution. This limitation arises mainly from two factors. First, the inner boundary condition in the Hydwind code is imposed at an optical depth $\tau = 2/3$, or equivalently at the corresponding mass density at that layer. As a consequence, once the inner boundary condition is fixed, the velocity gradient $dv/dr$ cannot smoothly pass through the critical point. Since the inner boundary conditions cannot be varied, the solution may therefore fail to satisfy the regularity and continuity conditions at the critical point. Although one could in principle adjust the boundary conditions to enforce these constraints, doing so within a stationary framework becomes impractical.

A time-dependent approach, however, offers an important advantage: the flow can dynamically adjust and relax toward a physically consistent solution, allowing the system to naturally approach the critical configuration. Nevertheless, the ZEUS-3D code does not always converge to a hydrodynamic solution. This issue was resolved by lowering the mass density at the inner boundary condition by one order of magnitude with respect to the photospheric density obtained from the Hydwind solution, where the photosphere is defined by imposing $\tau = 2/3$. Once a solution was obtained, we identified, from an appropriate Tlusty/Kurucz \citep{kurucz, hubeny1995} atmospheric model matching the stellar parameters, the radius at which this density is reached (typically around $1.01\,\rm{R}_*$), and then reran the ZEUS-3D simulations imposing the boundary condition at that corresponding photospheric layer. In this way, we ensure that the photospheric density stratification and the wind are smoothly connected.

\subsection{Comparison of line profiles computed with the traditional $\beta$-law and the hydrodynamic models}

\begin{figure*}
    \centering
    \includegraphics[width=0.9\linewidth]{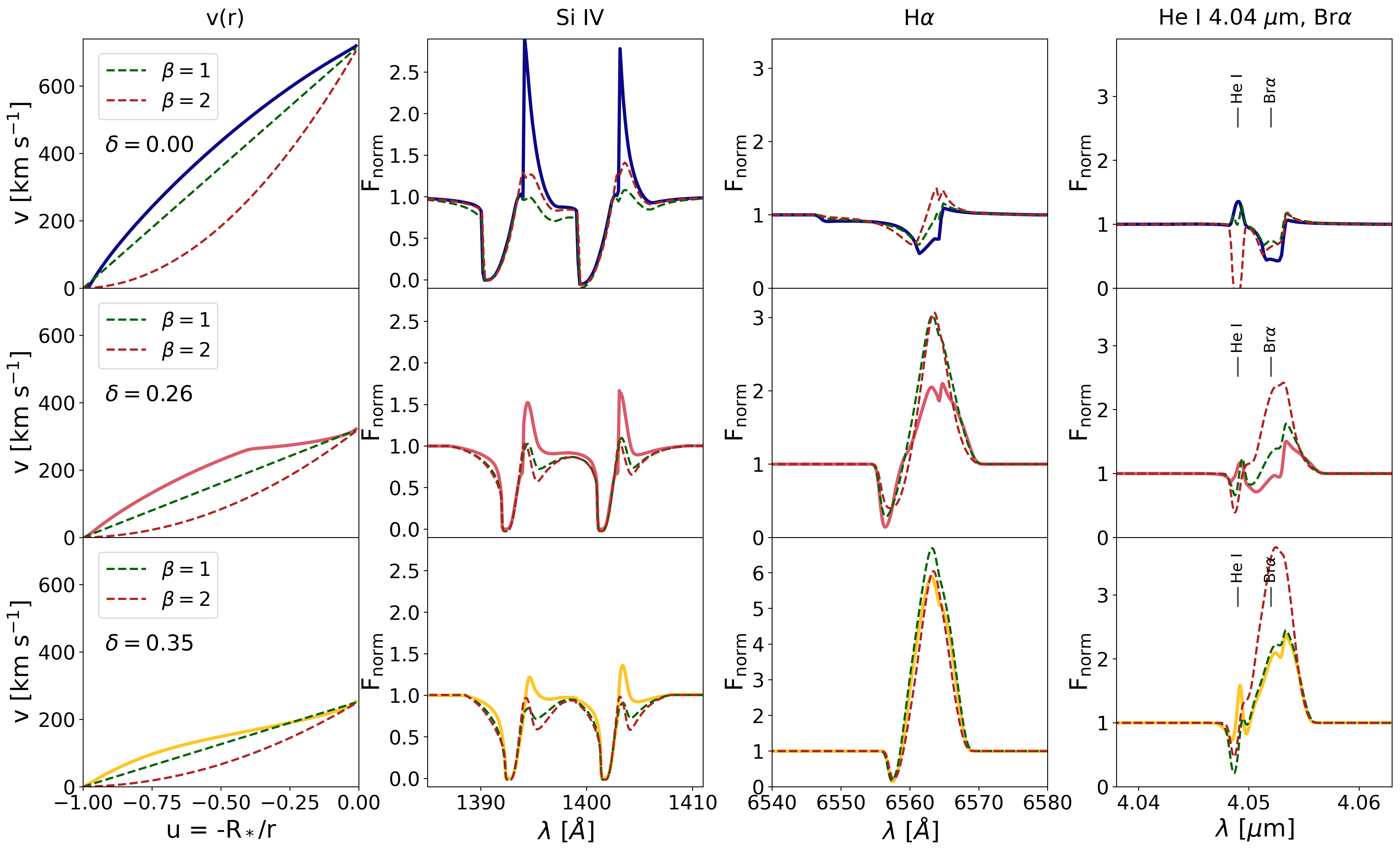}
    \caption{Comparison between synthetic line profiles computed using the hydrodynamic solutions obtained with ZEUS-3D (using the T19 model) and those derived from standard $\beta$-type velocity laws. The colour coding of the hydrodynamic solutions is the same as in Fig.~\ref{fig:profiles}: fast solution ($\delta = 0$), transition solution ($\delta = 0.26$), and $\delta$-slow solution ($\delta = 0.35$). Dashed lines correspond to the profiles computed adopting $\beta$ laws with $\beta = 1$ (green) and $\beta = 2$ (red). The columns show, from left to right, the hydrodynamic velocity structure, followed by the synthetic profiles for \ion{Si}{iv}, H$\alpha$, and the \ion{He}{i}~+~Br$\alpha$ blend around 4.05~$\mu$m.
}
    \label{fig:comp-beta}
\end{figure*}

Figure~\ref{fig:comp-beta} shows the line profiles computed with the hydrodynamic solutions (obtained with ZEUS-3D) together with those calculated from $\beta$-laws with $\beta = 1$ and $\beta = 2$. In all cases, the synthetic spectra were computed by adopting the same mass-loss rate, terminal velocity, and initial velocity as those of the corresponding hydrodynamic model.

Reproducing the H$\alpha$ line profiles typically observed in the winds of B supergiants generally requires $\beta$ values in the range $\sim 1.5 - 2.5$ \citep[e.g.,][]{2008markovapuls,haucke2018}. However, such velocity laws deviate significantly from any hydrodynamic wind solution. Nevertheless, it is known that a degeneracy exists in reproducing the overall shape of line profiles when different velocity laws are adopted \citep{2024venero}. In other words, both the traditional $\beta$-law and hydrodynamic solutions can produce similar spectral signatures under certain conditions. Therefore, comparing the synthetic line profiles obtained from both velocity prescriptions provides a useful way to evaluate their impact on the emergent spectra and to assess which hydrodynamic regimes can account for the large $\beta$ values inferred from observations.

It is worth noting that the H$\alpha$ profiles computed with a large $\beta$ value tend to resemble the profile obtained from the $\delta$-slow hydrodynamic solution. This may have implications for previous studies based on $\beta$-laws, as similar H$\alpha$ profiles could be associated with different underlying wind solutions, but with the same mass-loss rate and terminal velocity, although a more systematic analysis would be required to assess the generality of this result. 

On the other hand, the infrared \ion{He}{i} emission at 4.049~$\mu$m
\citep[typically seen in BSG stars, e.g.,][]{cidale2023} can, in some cases, only be reproduced by adopting $\beta$-laws with a low $\beta$ value, while the hydrodynamic solutions are able to reproduce this feature more naturally, indicating a higher sensitivity of this diagnostic to the adopted velocity structure. At the same time, the resulting emission predicted by each velocity law shows a different qualitative behaviour depending on the spectral range considered. These results reflect the interplay between velocity structure and mass-loss rate in shaping the emergent line profiles, and the need to combine diagnostics from different spectral ranges to better describe the wind dynamics.

\subsection{The ionisation structure of the wind}\label{subsection:5.3}

There is growing evidence for anomalous ionisation conditions in the winds of B supergiants. One example is the bi-stability jump, where both the mass-loss rate and the terminal velocity change for stars with effective temperatures around $21\,000$~K \citep{1995lamers,2008markovapuls}. Additional evidence comes from the superionisation effect observed in the UV, which has led to the suggestion of hot regions in the wind and the presence of X-ray emission \citep{1981cassinelli,2016krtickakubat}. At the same time, optical spectra show lines from relatively low ionisation stages, and wind-clumping models are often required to explain the resulting opacity structure to fit the line features observed in different spectral ranges.

At present, there is no self-consistent model that demonstrates how the ionisation parameter $\delta$ varies with spectral type, density distribution, and radial distance in the wind. Nevertheless, the fact that the empirical $\beta$-law required to reproduce the observed line profiles resembles a slow wind solution (i.e., associated with relatively large $\delta$ values) suggests that the ionisation structure of the wind may play an important role. 

Although most applications of the standard m-CAK formalism adopt relatively low $\delta$ values, larger values are not excluded on physical grounds. Analytical considerations indicate that $\delta \gtrsim 1/3$ can arise in media where neutral hydrogen behaves as a trace element, implying strong ionisation variations throughout the wind \citep{2000puls+}. In addition, approximate non-LTE calculations suggest that $\delta$ may approach unity in low-density winds or in low-metallicity environments \citep{2002kudritzki}. In this context, the range of $\delta$ values explored in this work can be regarded as physically plausible, particularly in view of the complex and not yet fully constrained ionisation structure of B supergiant winds. This indicates that further work in this direction is needed, especially using models that relax the assumption of isothermal winds.

\section{Conclusions}\label{section:6}

In this work, we modelled radiation-driven winds of massive stars within the m-CAK framework, focusing on the transition region between the fast and $\delta$-slow regimes. Using time-dependent ZEUS-3D simulations, we demonstrated that B supergiant winds admit a continuous sequence of stationary hydrodynamic solutions that extends beyond the traditionally recognised regimes. The region previously thought to lack stationary solutions is now fully resolved: we uncovered a well-defined set of intermediate solutions that smoothly connects the fast and $\delta$-slow regimes. Several of these solutions develop a distinct kink in the velocity law, whose location shifts outward as $\delta$ decreases and becomes steeper with increasing rotation. These results show that the apparent discontinuities obtained with stationary approaches do not represent a physical gap in the parameter space, but rather the limitations of earlier methods.

We also computed synthetic line profiles for H, \ion{He}{i}, and \ion{Si}{iv} by solving the radiative transfer in the comoving frame for different hydrodynamic solutions to illustrate the effect of the wind structure. A key contribution of this work is the inclusion of a detailed H, \ion{He}{i} and \ion{Si}{iv} atomic model within a moving-medium radiative transfer framework, enabling self-consistent predictions of ultraviolet and optical lines together with a comprehensive set of infrared transitions.
 
The resulting profiles reflect the underlying hydrodynamic structure: the solutions in the transition region produce intermediate line morphologies between the fast and $\delta$-slow winds, providing direct spectroscopic evidence for the continuity of the hydrodynamic sequence. To further assess the impact of the adopted velocity structure, we compared these profiles with those computed using standard $\beta$-type velocity laws. In the model explored, H$\alpha$ profiles computed using standard $\beta$-type velocity laws, with $\beta > 1$, closely resemble the profile obtained from the $\delta$-slow hydrodynamic solution. This suggests that similar H$\alpha$ signatures may arise from different underlying wind structures, when the mass-loss rate and terminal velocity are the same, with potential implications for previous studies based on $\beta$-law prescriptions. However, a more systematic exploration of the parameter space is required to assess its general validity.

These results establish a unified picture in which the wind properties of B supergiants evolve smoothly with the ionisation parameter $\delta$, without invoking additional physical mechanisms to reproduce the observed line morphologies. The existence of this continuous sequence opens the door to a more refined interpretation of spectroscopic diagnostics across the UV–optical–IR range, suggesting that the hydrodynamic state of B-supergiant winds can be robustly constrained through simultaneous multiwavelength observations. Furthermore, perturbations in the wind ionisation alter the line force and may trigger transitions between different hydrodynamic regimes. Such ionisation-driven regime changes offer a plausible explanation for structured and variable winds, including the emergence of large-scale features such as discrete absorption components. A systematic investigation of these effects will be the subject of future work.

\begin{acknowledgements}
      We thank the referee, Nicolas Moens, for the careful reading and constructive comments, which helped to improve the clarity and quality of this work. MCF, ROJV, and LSC acknowledge financial support from CONICET (PIP 11220200101337CO) and from the Universidad Nacional de La Plata through the Programa de Incentivos (grant 11/G192). ROJV also acknowledges support from grant 11/G193. IA and MC thank the FONDECYT projects 1230131 and 1261498 for their support.
      This work has been partially co-funded by the European Union, Project 101183150 - OCEANS.
\end{acknowledgements}

\bibliographystyle{aa}
\bibliography{cites}

\clearpage
\onecolumn
\begin{appendix} 
\section{Terminal velocity and mass-loss rates for the T19 model ($\Omega = 0.2$)}

{\renewcommand{\arraystretch}{1.2}
\begin{table*}[ht!]
\caption{Stellar wind parameters for the solutions found in the fast and $\delta$-slow regimes, and in the gap region for the T19 model with $\Omega = 0.2$, computed using ZEUS-3D and Hydwind codes.}           
\label{table:1}      
\centering                         
\begin{tabular}{c c c c c c}        
\hline\hline                 
\multirow{3}{*}{Regime} & \multirow{3}{*}{$\delta$} & \multicolumn{2}{c}{Hydwind} & \multicolumn{2}{c}{ZEUS-3D} \\
\cmidrule(lr){3-4} \cmidrule(lr){5-6}
& & $v_{\infty}$ & $\dot{M}$ & $v_{\infty}$ & $\dot{M}$ \\    
& & [km s$^{-1}$] & [10$^{-6}$ M$_{\odot}$ yr$^{-1}$] & [km s$^{-1}$] & [10$^{-6}$ M$_{\odot}$ yr$^{-1}$] \\   
\hline 
   \multirow{21}{*}{Fast} & 0.00 & 703.4 & 1.261 & 708.5 & 1.092 \\ 
   & 0.01 & 683.3 & 1.281  & 688.4 & 1.106 \\
   & 0.02 & 663.9 & 1.301  & 669.0 & 1.120 \\
   & 0.03 & 645.2 & 1.322  & 650.3 & 1.134 \\
   & 0.04 & 627.1 & 1.344  & 632.2 & 1.149 \\
   & 0.05 & 609.6 & 1.368  & 614.7 & 1.164 \\
   & 0.06 & 592.8 & 1.392  & 597.8 & 1.181 \\
   & 0.07 & 576.5 & 1.417  & 581.4 & 1.197 \\
   & 0.08 & 560.6 & 1.444  & 565.5 & 1.215 \\
   & 0.09 & 545.3 & 1.472  & 550.1 & 1.234 \\
   & 0.10 & 530.4 & 1.502  & 536.7 & 1.249 \\
   & 0.11 & 515.9 & 1.534  & 536.7 & 1.269 \\
   & 0.12 & 501.7 & 1.567  & 507.8 & 1.291 \\
   & 0.13 & 488.0 & 1.603  & 493.9 & 1.313 \\
   & 0.14 & 474.5 & 1.641  & 480.3 & 1.338 \\
   & 0.15 & 461.3 & 1.682  & 467.0 & 1.363 \\
   & 0.16 & 448.4 & 1.727  & 453.9 & 1.391 \\
   & 0.17 & 435.6 & 1.774  & 443.3 & 1.411 \\
   & 0.18 & 423.1 & 1.826  & 430.5 & 1.442 \\
   & 0.19 & 410.6 & 1.882  & 417.9 & 1.476 \\
   & 0.20 & 398.1 & 1.943  & 405.3 & 1.512 \\
   & 0.21 & 385.4 & 2.010  & 392.5 & 1.551 \\
   \hline
   \multirow{6}{*}{Gap} & 0.22 & \dots & \dots & 379.0 & 1.594 \\
   & 0.23 & \dots & \dots & 359.2 & 1.635 \\
   & 0.24 & \dots & \dots & 339.0 & 1.674 \\
   & 0.25 & \dots & \dots & 323.9 & 1.708 \\
   & 0.26 & \dots & \dots & 308.3 & 1.761 \\
   & 0.27 & \dots & \dots & 294.5 & 1.827 \\
   & 0.28 & \dots & \dots & 283.1 & 1.901 \\
   \hline
   \multirow{11}{*}{$\delta_{\textnormal{slow}}$} & 0.29 & 262.9 & 2.882 & 274.7 & 1.974 \\
   & 0.30 & 258.8 & 3.017 & 267.1 & 2.055 \\
   & 0.31 & 255.2 & 3.169 & 262.0 & 2.125 \\
   & 0.32 & 252.0 & 3.348 & 258.0 & 2.200 \\
   & 0.33 & 248.9 & 3.561 & 254.5 & 2.284 \\
   & 0.34 & 246.0 & 3.820 & 251.3 & 2.383 \\
   & 0.35 & 243.2 & 4.138 & 248.3 & 2.500 \\
   & 0.36 & 240.6 & 4.537 & 245.5 & 2.643 \\
   & 0.37 & 238.1 & 5.045 & 242.9 & 2.820 \\
   & 0.38 & 235.9 & 5.704 & 241.0 & 3.012 \\
   & 0.39 & 234.0 & 6.569 & 239.7 & 3.237 \\
   & 0.40 & 232.7 & 7.700 & 239.0 & 3.483 \\
\hline                                   
\end{tabular}
\end{table*}}

\end{appendix}

\end{document}